# Origin of Coda Waves: Earthquake Source Resonance

Yinbin Liu
Burnaby, British Columbia, Canada
(Email: yliu@eoas.ubc.ca)

**Abstract**
Coda in local earthquake exhibits resonance-like wave behaviour where the coda emerges as long-duration small-amplitude vibration with selective frequency, slow temporal decay, and uniform spatial energy distribution around the earthquake source. Coda is thought to be the incoherent waves scattered from random small-scale heterogeneity in the earth's lithosphere. Here I show that the coda in local earthquake is primarily attributed to the natural resonance in strong small-scale heterogeneity around the earthquake's hypocenter through seismic wave field modeling for 1D heterogeneity. The natural resonance is evolved from the low frequency resonance (LFR) and is an emergent phenomenon that occurs in steady state regime. Effects of heterogeneity on the natural resonance are similar to those of LFR.

**Introduction**

When you strike a bell, the kinetic energy is converted into sound energy within the bell. Strong acoustic impedance contrast between the bell and the surrounding air causes the multiple scattering of wave that forms trapped-energy resonance within the bell. The resonance energy continuously leaks into the surrounding medium, disperses uniformly around the bell, and shows slow temporal decay and selective frequency feature. The resonance frequency is inversely proportional to the size of the bell and is independent of the location where you strike the bell or where you hear the ringing sound. Coda in local earthquakes exhibits the similar characteristics of a bell ringing, i.e., the uniform spatial energy distribution around the earthquake source, the same frequency contents at all recording stations, and slow temporal decay oscillation (*1*-*5*).

Coda, which is traditionally defined as the tail of a seismogram, is usually thought to originate from the high-order scattered waves from numerous heterogeneities in the earth's lithosphere but never related to the Earthquake's hypocenter (*1*-*5*). Earthquake source zone is always associated with strong small-scale heterogeneity, for example, gas-and-fluid-related subduction zone (*6*). Strong small-scale heterogeneity tends to trap seismic energy that may result in complex temporal evolution of wave packet. Based on scattered wave field modeling in 1D heterogeneity, this study shows that the multiple scattering of seismic waves in strong small-scale heterogeneity may cause many-body system natural resonance in steady state regime, which is a kind of emergence phenomenon evolved from LFR in transient regime (*7*). The natural resonance around the earthquake's hypocenter exhibits features similar to a bell ringing and provides a physical interpretation on coda in the local earthquake.

**Coda**

Figure 1 shows original seismograms (vertical) and their corresponding spectra, which include the influence of random noise such as winds and tides, recorded at the same station (MGB) for two different magnitude local earthquakes in Vancouver Island.

The earthquakes have very close earthquake's hypocenter (about 10 km) and the distance from the station to the epicentre is about 310 km (Fig. 1B).

Figure 1A is the comparison of the seismograms of the two earthquakes. The blue and the dark red lines stand for a M 4.8 earthquake (48.68N128.94W) and a M 5.1 earthquake (48.68N128.89W), respectively. The amplitude for the M 4.8 earthquake has been amplified 4 times for comparison, Fig. 1C to 1F are a stretched-out version of Fig. 1A. It can be seen that the seismograms are composed of three groups of wavetrains in the time domain: the early-arrival high-frequency wave component superposed on a low-frequency wave component (Fig. 1C); the following low-frequency large-amplitude main wave component with increasing instantaneous frequency (Fig. 1D); and the late-arrival coda (Figs. 1E and 1F). The low-frequency main wave component, which is conventionally thought to be Rayleigh-type surface wave, has a travel-time about 80 s (Fig. 1D) and a dominant frequency about 0.05 Hz (Fig. 1H); thus the direct propagation distance from the source to the station is only about four seismic wavelengths. The late-arrival coda goes on for more than 15 times the traveling time (long traveling paths) of the direct wave and reveals very slow temporal decay (strong nonlinear interaction). These observations manifest that the coda is associated with a kind of very high order multiple scattering from strong heterogeneity.

From the viewpoint of hierarchical structure, this study views the seismograms as a superposition of an early-arrival high-frequency wave component (the behaviour of Sommerfeld precursor field) and a low-frequency wave component (the behaviour of Brillouin precursor field). The latter is composed of an early-arrival low-frequency wave component, a low-frequency large-amplitude main wave component, and coda. This feature of seismograms is similar to the long period event (a high-frequency onset superposing on a low-frequency background) in volcanic seismology (*8*) but occurs at a relatively longer time scale. Figs. 1C to 1F show excellent agreement between the two seismograms for the low-frequency wave component for the first about 700 s (a new earthquake appears after the first 700 s). The agreement indicates that the low-frequency wave component is associated with some kinds of linear wave propagation and scattering effects through the complex heterogeneous medium system from the source to the station (nonlinearity will lead to intensity-dependent wave packet evolution).

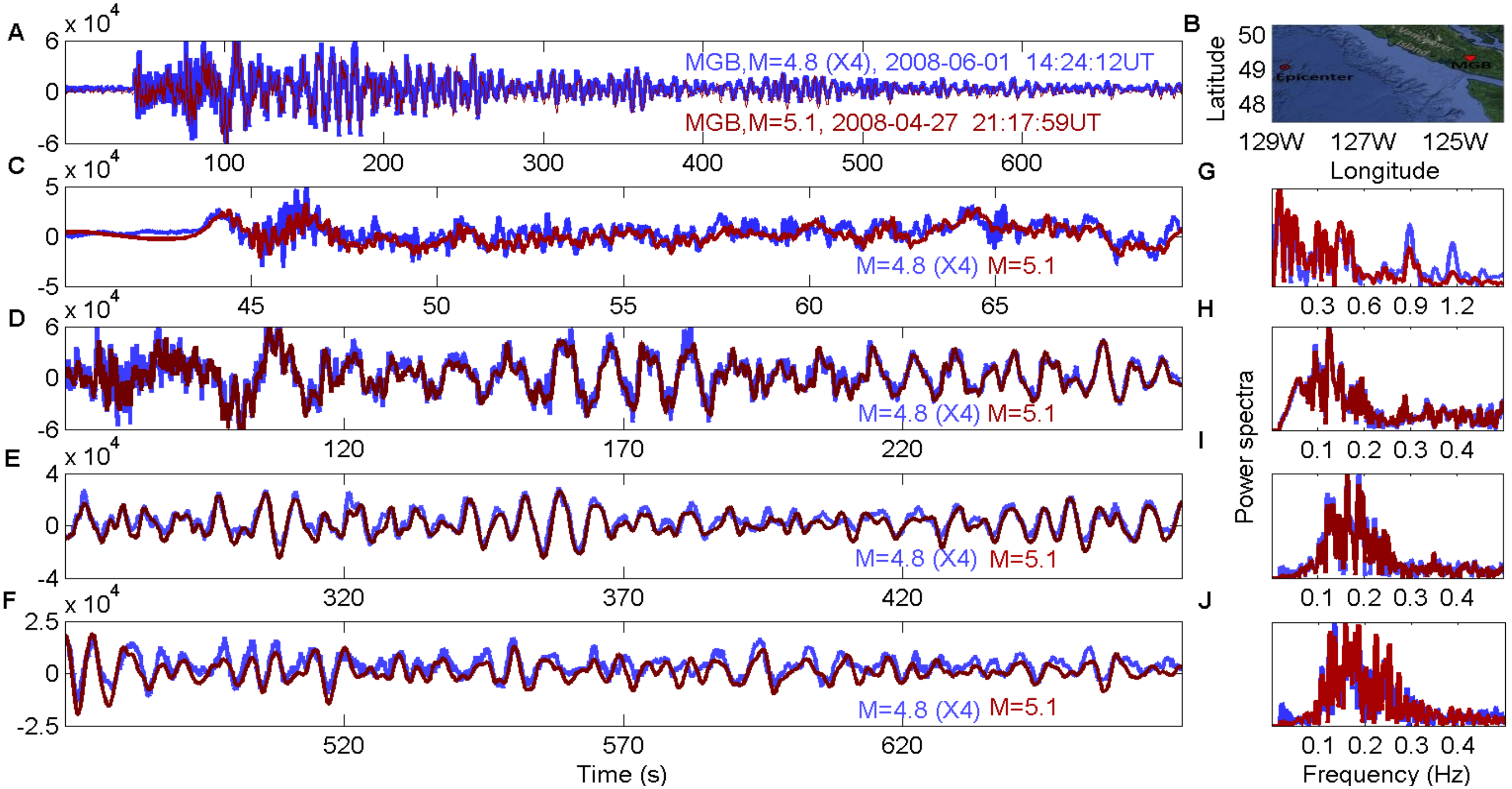

**Fig. 1.** Comparisons of seismograms at station MGB and the same earthquake's focus location (10.0 km) from two local earthquakes in Vancouver Island. (**A)** Seismograms for a 4.8-magntitude earthquake (48.68N128.94W) on 1 June 2008, which is amplified by a factor 4, and a 5.1-magntitude earthquake (48.68N128.89W) on 27 April 2008. (**B**) Schematic map of earthquake epicenter station MGB. (**C** – **F**) Comparison of the seismograms for the two earthquakes. (**G** – **J**) Normalized power spectra.

The normalized power spectra in Figs. 1G to 1J exhibit a complex structure featuring a series of peaks. Fig. 1G shows that there are two high-frequency peaks at about 0.9 Hz and 1.2 Hz, their relative intensities are a little larger for the M 4.8 earthquake than for the M 5.1 earthquake as expressed in the early-arrival high-frequency wave component in Fig. 1C. This amplitude discrepancy is likely due to the influence of different excitation or release of stress accumulated during tectonic deformation. Fig. 1G also shows that there are several low-frequency peaks associated with the early-arrival low-frequency wave component, the first peak is their maximum peak at about 0.05 Hz. As propagation time increases, the low-frequency wave component evolves into main wave component and coda (Figs. 1D to 1F). The dominant frequency of the main wave component is corresponding to the first peak at about 0.05 Hz (Fig. 1H), and the maximum peak frequency of the coda is about 0.12 Hz in Fig. 1H, 0.16 Hz in Fig. 1I, and 0.18 Hz in Fig. 1J, respectively. The instantaneous frequencies of the coda tend to increase monotonically and have stable frequencies in steady state regime (*2*). This implicates that the coda might not be attributable to intrinsic absorption in the lithosphere, which brings about instantaneous frequency decreasing over duration because intrinsic absorption tends to attenuate higher frequencies faster than lower frequencies. The dominant frequency of the first cyclic low-frequency main wave component is about 20 times lower than that of the high-frequency wave component and about 3.7 times lower than that of the coda. Another feature of the coda is the non-uniform temporal energy distribution, i.e., the wave packet intensity of the late-arrival coda might be higher than that of the early-arrival coda as seen between 370 s to 490 s in Figs. 1E and 1F. These dynamic coda scattering properties cannot be described by diffusion approximation (*9*) and the so-called energy equipartition (*10*) based on radiative transfer theory.

**Natural Resonance in Strong 1D Heterogeneity**

Delta propagator approach (*11*) is employed to numerically study the dynamic coda wave scattering properties in strong 1D heterogeneity, which is composed of period two constituent layered units embedded between two fluid half-spaces. Two small-scale heterogeneities are constructed by choosing a total thickness $D = D_1 + D_2 = 208$ m ( $D_1 = 68\,\mathrm{m}$ and $D_2 = 140\,\mathrm{m}$ ) and two lattice constants $d = d_1 + d_2 = 6.5\,\mathrm{m}$ (64 layers, $d_1 = 2.125\,\mathrm{m}$ ) and $d$ = 3.25 m (128 layers, $d_1 = 1.0625\,\mathrm{m}$ ). The physical properties of constituent units are the same as Liu (*7*). The incident pulse is a single cycle pulse (olive in Figs. 2 to 5, with scaled-down amplitude) with a dominant frequency of $f_s = 172$ Hz (dash olive in Figs. 2 to 5).

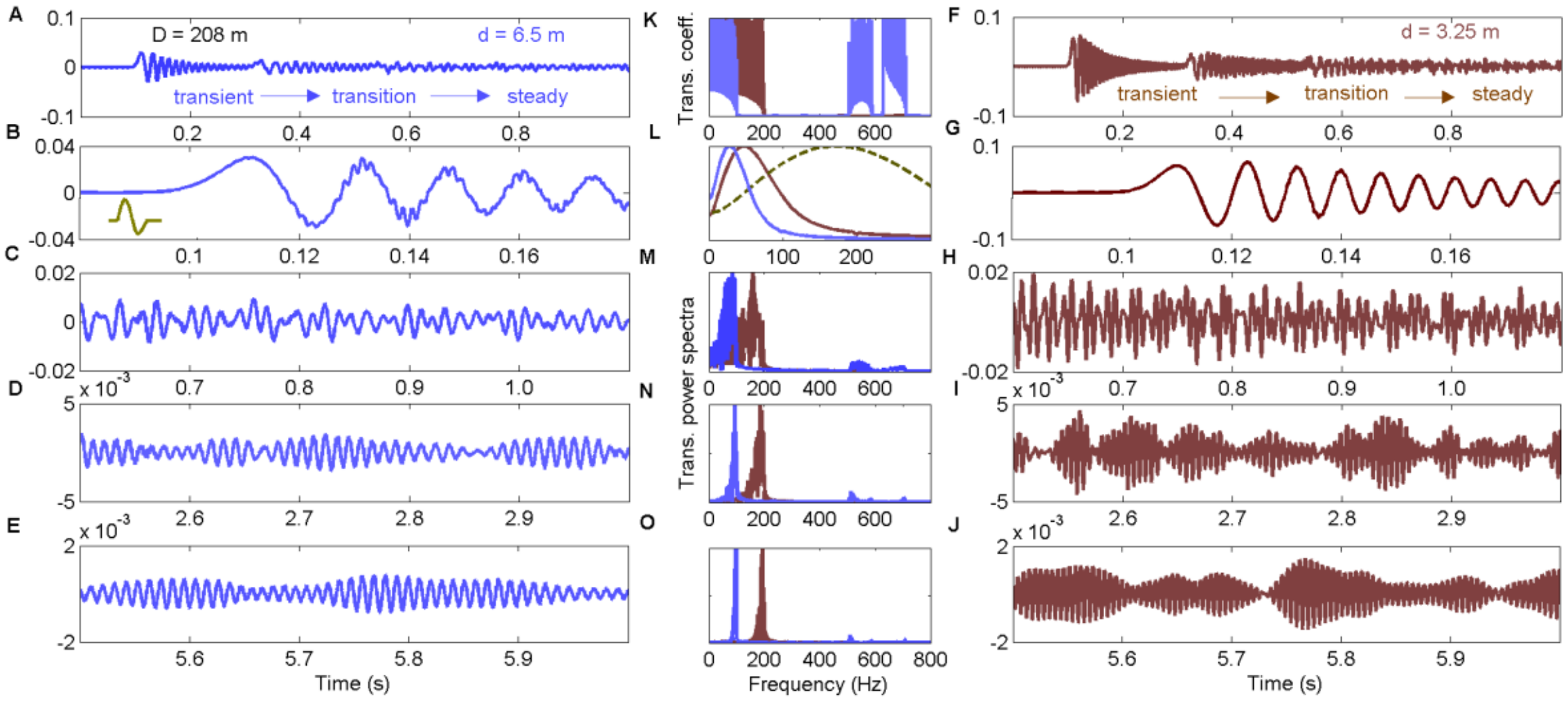

**Fig. 2.** Scale-dependent natural resonance. Plastic/steel heterogeneity with total thickness $D$ = 208 m and lattice constants $d$ = 6.5 m or $d_1 = 2.125\text{m}$ (blue) and $d$ = 3.25 m or $d_1 = 1.0625\text{m}$ (dark red). (**A – E**) Normal transmission wave fields for $d_1 = 2.125\text{m}$. (**F – J**) Normal transmission wave fields for $d_1 = 1.0625\text{m}$. (**K**) Transmission coefficients. (**L – O**) Normalized power spectra.

Figures 2A and 2F show the normal transmission wave fields for plastic/steel heterogeneity with heterogeneous scale $d_1 = 2.125\text{m}$ (blue) and $d_1 = 1.0625\text{m}$ (dark red), respectively. Figs. 2B to 2E and Figs. 2G to 2J are the corresponding stretched-out versions of Figs. 2A and 2F. The wave fields in the time domain can be roughly separated into transient regime, transition regime, and steady state regime. The arrivals of the first and second wavetrain groups are at about 0.1 s (the first half-space transmission) and 0.32 s (the second half-space transmission). Fig. 2K is the transmission coefficients and Figs. 2L to 2O are the normalized power spectra. The larger the lattice constant, the lower the upper stopband corner frequency occurs.

The low-frequency wave component in Figs. 2B and 2G is associated with low-frequency resonance (LFR) (*7*) in transient regime and evolves into very slow temporal decay oscillations in transition and steady state regimes with a non-uniform temporal energy distribution. The wave fields in transition regime (Figs. 2C and 2H) exhibit more complex structures than those in steady state regime (Figs. 2D, 2E, 2I, and 2J); their corresponding power spectra (Figs. 2M to 2O) show that the instantaneous frequencies tend to increase monotonically and have stable frequencies in steady state regime. The wave field in steady state regime is associated with very high-order multiple scattering and the number of scattering orders can be estimated by the arrival time of wave field and the traveling time of constituent units. For example, the ray traveling time passed through an individual plastic or steel layer is about 0.4 ms for $d_1 = 1.0625\text{m}$, thus the wave packet arrived at about 2.6 s in Fig. 2I has undergone up to about 6500th order scattering. The longer the scattered time or scattered path, the simpler the structure of wave field, or the narrower the corresponding spectrum peak. This is because the resonance characterization of the system will dominate over the wave propagation and exhibits selective frequency feature in steady state regime. The dominant frequencies of power spectra of wave fields in steady state regime are about 98 Hz for $d_1 = 2.125\,\mathrm{m}$ and about 198 Hz for $d_1 = 1.0625\,\mathrm{m}$ (Figs. 2N and 2O), the frequencies are inversely proportional to heterogeneous scale. I call this phenomenon natural resonance in strong small-scale heterogeneity, which is a kind of coherent scattering enhancement or emergence phenomenon in steady state regime. LFR in transient regime and natural resonance in steady state regime are two distinct types of evolutionary processes. Figs. 2M to 2O also show a series of high frequency small-amplitude peaks (less than 11% at about 510 Hz, 585 Hz, 675 Hz, and 710 Hz for $d_1 = 2.125\,\mathrm{m}$) that mainly associated with individual plastic or steel layer resonance. Note that the frequencies of natural resonance in Fig 2 are about 3.6 times higher than those of the corresponding LFR (*7*) and about 6 times lower than those of the corresponding individual layer resonance (the fundamental resonance frequency of an individual layer is 585 Hz or 1170 Hz for the plastic and 633 Hz or 1265 Hz for the steel for $d_1 = 2.125\,\mathrm{m}$ or $d_1 = 1.0625\,\mathrm{m}$, respectively). This modeling also demonstrates that the wave fields in steady state regime exhibit a uniform spatial energy distribution, i.e., the reflection fields and transmission fields have the same spectrum and strength characteristics. All those features of the natural resonance are similar to those of the observed coda in Fig. 1, i.e., selective frequency, uniform spatial distribution energy, and non-uniform temporal decay oscillation with increasing instantaneous frequency.

Figure 3 has the same expressions as Fig. 2 except shows the transmission wave fields and their corresponding spectra are for shale/gas I heterogeneity (Figs. 3A to 3E) and shale/gas II heterogeneity (Figs. 3F to 3J). The wave fields in Figs. 3B and 3G show a superposition of the early-arrival high-frequency small-amplitude wave component and the low-frequency resonance. The former is associated with the resonance of the individual shale or gas layer and the latter will evolve into the natural resonance of the system in steady state region. The energy of the natural resonance is proportional to the impedance contrast of constituent units. The larger the impedance contrast, the stronger the energy. The dominant frequency of natural resonance of the system is about 98 Hz for plastic/steel heterogeneity ($d_1 = 2.125\,\mathrm{m}$ in Fig. 2), 40 Hz for shale/gas I heterogeneity, and 22 Hz for shale/gas II heterogeneity. The frequency of the natural resonance decreases with increasing impedance contrast. Figs. 3M to 3O also show a series of high frequency peaks that are mainly associated with individual gas or shale layer resonance at about 220 Hz, 240 Hz, 310 Hz, 330 Hz, 465 Hz, and 475 Hz for shale/gas I heterogeneity and about 165 Hz, 300 Hz, 315 Hz, 330 Hz, 345 Hz, and 495 Hz for shale/gas II heterogeneity. These peaks exhibit complex spectral structure and temporal evolution and are strongly dependent on the spatial symmetry of the constituent units. For instance, a very small peak (about 4%) at about 165 Hz in Fig. 3M evolves into a maximum peak in Fig. 3N or 3O and the high frequency peaks disappear with the spatial symmetry breaking as shown in Figs. 3C and 3M (dark green) for a 1% root-mean-square (RMS) velocity fluctuation for shale/gas I heterogeneity. The frequency of natural resonance is about 3.6 times (or 3.7 times) higher than that of LFR and about 5.9 times (or 7.5 times) lower than the resonance frequency of an individual gas layer for shale/gas I heterogeneity (or shale/gas II heterogeneity). This modeling also demonstrates that the gas proportion has little influence on the frequency of natural resonance.

If the early-arrival high-frequency wave component in transient regime and the wave field in the transition regime are viewed as a kind of disorder or non-equilibrium process and LFR in transient regime and the natural resonance in steady state regime as a kind of order or equilibrium process, the long-time dynamic wave scattering undergoes the evolutions from disorder-to-order-to-disorder-to-order processes that occur at different hierarchical structures. This indicates that a disordered state in a low-order hierarchical structure can evolve to an ordered state in a high-order hierarchical structure through coherent wave scattering or self-organization in an open system.

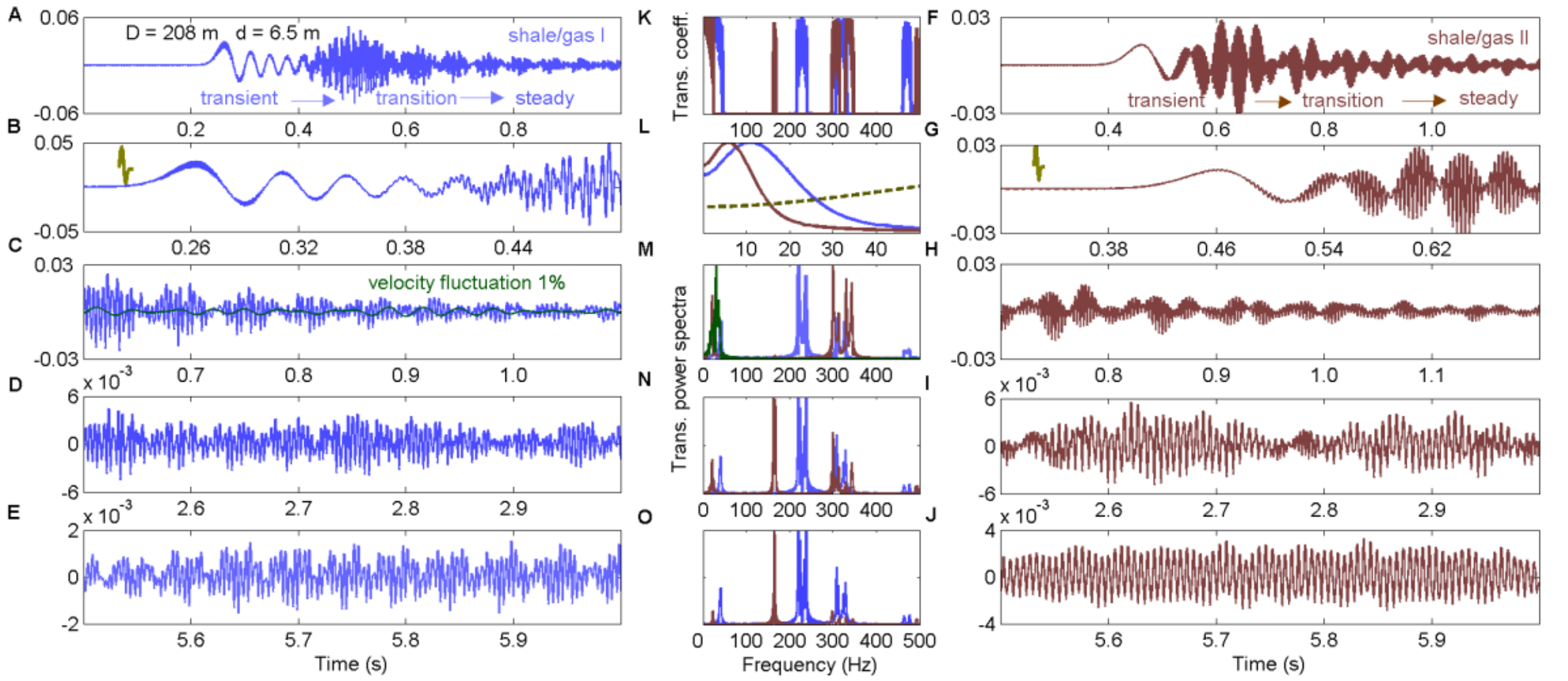

**Fig. 3.** Contrast-dependent natural resonance. The same as Fig. 2 except for shale/gas I and shale/gas II heterogeneities with $d_1 = 2.125\,\text{m}$.

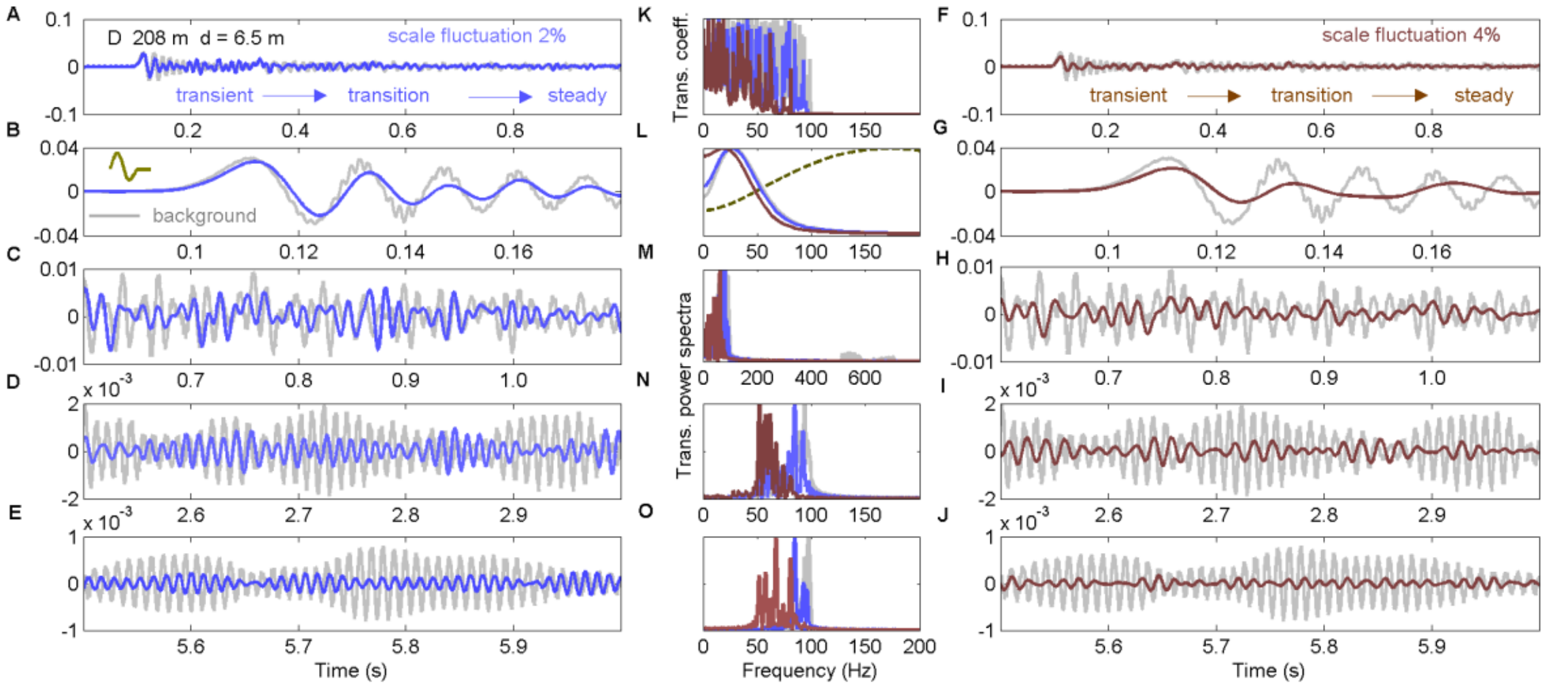

**Fig. 4.** Effect of random scale fluctuation on natural resonance. Plastic/steel heterogeneity with a lattice constant $d$ = 6.5 m ( $d_1$ = 2.125 m), total thickness $D$ = 208 m, and different scale fluctuations. (**A** – **J**) Normal transmission wave fields for RMS scale fluctuations $\delta d/d$ = 2% (blue) and 4% (dark red). (**K**) Transmission coefficients. (**L** – **O**) Normalized power spectra.

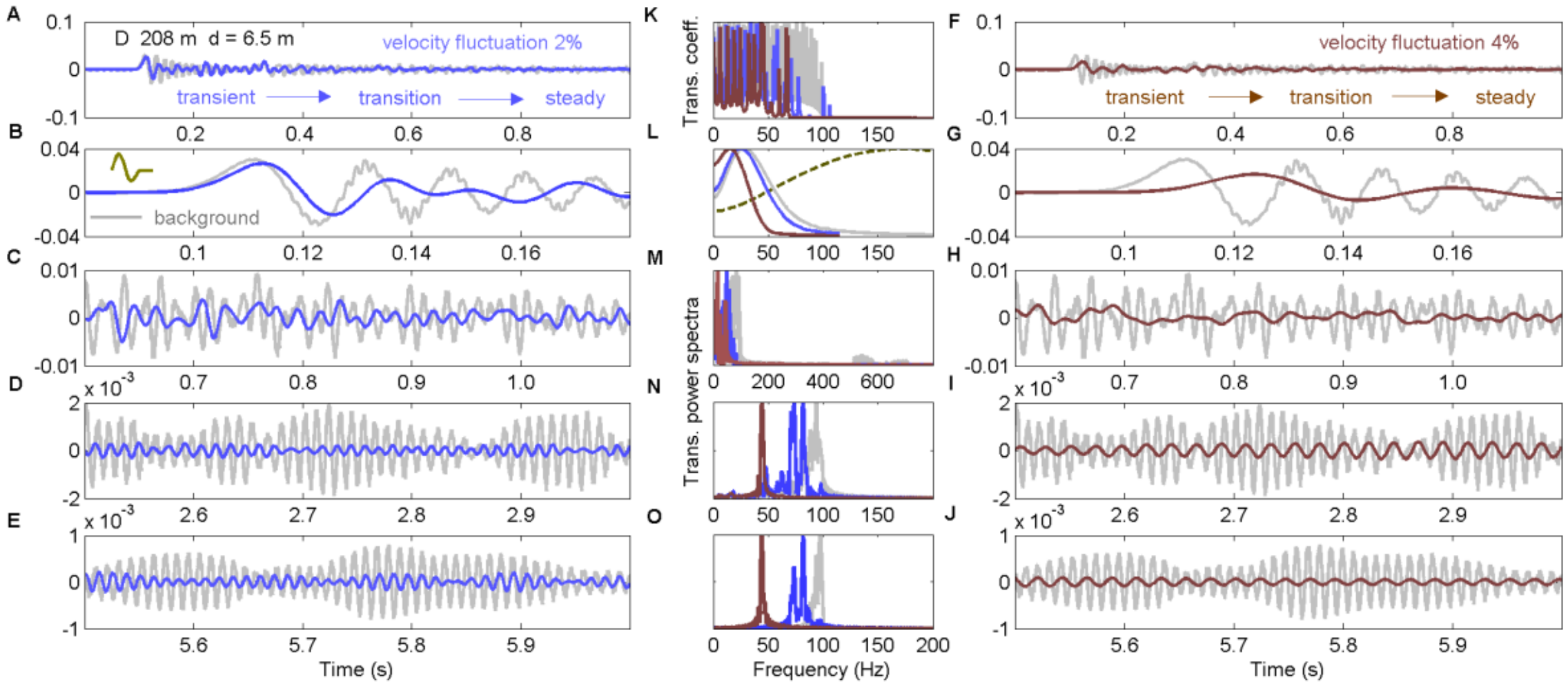

**Fig. 5.** Effect of random velocity fluctuation on natural resonance. The same as Fig. 4 except for RMS velocity fluctuations $\delta v/v$ = 2% (blue) and 4% (dark red).

Figures 4 and 5 show the influences of random scale (Fig. 4) and velocity (Fig. 5) fluctuations of plastic/steel heterogeneity on natural resonance. The blue and dark red stand for 2% and 4% the root-mean-square (RMS) scale and velocity fluctuations (the grey for the background), respectively. An increase in the scale and velocity fluctuations means a decrease in the spatial symmetry of small-scale heterogeneity. The high frequency small-amplitude peaks associated with individual layer resonance will disappear with the spatial symmetry breaking as shown for scale and velocity fluctuations in Figs. 4M and 5M. The frequency of natural resonance is about 84.5 Hz for $\delta d/d$ = 2% and 67 Hz for $\delta d/d$ = 4% for scale fluctuations; and is about 81.5 Hz for $\delta v/v$ = 2% and 44 Hz for $\delta v/v$ = 4% for velocity fluctuations (98 Hz for the background). The natural resonance is a little more sensitive to the velocity than the scale fluctuations. The frequency of natural resonance decreases with increasing random heterogeneous scale and velocity fluctuations; and its energy also decreases with increasing scale and velocity fluctuations (Figs. 4 and 5). These features suggest that the frequency and strength of natural resonance will decrease with the lowering of the degree of spatial symmetry of small-scale heterogeneity.

## Discussions

Multiple scattering of seismic waves in strong small-scale 3D heterogeneity around the earthquake’s hypocenter will cause the dynamic emergence phenomenon that occurs as LFR in transient region and natural resonance in steady state region similar to strong small-scale 1D heterogeneity. The natural resonance will show the feature of a uniform spatial energy distribution around earthquake’s hypocenter and exhibits different resonance frequencies from region to region. Thus it seems that the natural resonance around the earthquake’s hypocenter is a more adequate interpretation on coda than the incoherent wave scattered from the earth’s lithosphere. Physically speaking, coda is arisen from very high-order multiple scattering and may include richly coherent scattering information. The procedure of high-order multiple scattering within strong heterogeneity is equivalent to physical multiple correlation (*12*) or time-reversal (*13*) in time and space in which the stronger gets relatively stronger and the weaker gets relatively weaker. After very high-order multiple scattering, the coherent scattering waves survive easier and form coherent coda waves, which exhibit wave packet evolution and selective frequency features in time domain. The coda wave in time domain and the wave localization in space domain often accompany the same multiple scattering phenomenon (*14*, *15*).

The coda waves from the earthquake source resonance might be further scattered by elsewhere in the lithosphere heterogeneity from the source to the receiver. This is similar to the ringing sound of a bell, where the wave is scattered or trapped by its surrounding media. The stronger the lithosphere heterogeneity, or the longer the scattered distance from the source to the receiver, the larger the influence of the lithosphere heterogeneity on the coda. For example, the frequency-wavenumber analysis shows that late coda waves at great distance are not the direct waves from the earthquake source but are the secondary waves generated by the different propagation paths or the local strong heterogeneity from the source to the receivers *(16)*. For a local earthquake, however, the influence is usually much smaller than that of the earthquake source resonance. This is because the initial rupture point of an earthquake is located in the earthquake source zone where natural resonance is easier to be excited than elsewhere in the earth's lithosphere.

If the coda in local earthquake mainly originates from the earthquake source resonance in steady state regime, the multiple scattering of waves in this kind of strong small-scale heterogeneity should also generate early-arrival high-frequency component and low-frequency resonance (Sommerfeld and Brillouin precursors) in transient regime (*7*). Thus the early-arrival high-frequency component and the low-frequency main wavetrain can be associated with earthquake's hypocenter (e.g., Fig. 1). The primary of Sommerfeld precursor itself is equivalent to the direct P-wave, and the formal solutions of LFR in 1D heterogeneity are equivalent to that of Rayleigh-type surface wave. However, only strong small-scale heterogeneity can generate LFR. The observed dynamic wave scattering features show that in volcanic earthquake the high-frequency component, low-frequency component with increasing instantaneous frequency, and coda are independent of the epicentral distance, azimuth, and time *(17)*, in glacial earthquake the location of the hypocenter from the short-period P-wave is in agreement with that from the long-period event in four near-source stations (*18*). These characteristics strongly support the assumptions that the low-frequency main wavetrain and coda in local earthquake are mainly related to LFR and natural resonance around the earthquake's hypocenter, respectively.

If the seismogram in local earthquake is mainly associated with earthquake source resonance, its dynamic scattering properties may provide a tremendous resource to monitor earthquake source characteristics. The high-frequency component reveals the individual constituent units and the low-frequency component reveals the ensembles of individual constituent units of small-scale heterogeneity. The waveform pattern and its frequency attribute (Fig. 1) are similar to those in strong small-scale 1D heterogeneity such as shale/gas heterogeneity. These features strongly support the perspective that the northern Cascadian subduction zone exhibits strong heterogeneity with high pore fluid pressure (*6*). Earthquake source zone is 3D heterogeneity and is likely to cause much more complex wave phenomena than those of 1D heterogeneity. The classic multiple scattering theory provides exact analytical series solutions (*19*) that may be developed to numerically study the collective behaviour in 2D and 3D many-body systems. Random matrix theory (RMT) studies the eigenvalue spacing distribution of response matrix for evaluating the symmetries and collectivities of the microscopic constituents (*20*). Two kinds of resonances from strong small-scale heterogeneity may provide new insights into RMT for evaluating microscopic or small-scale heterogeneity.

**Acknowledgements**

I thank my wife, Xiaoping Dai and my daughter, Wenbo Elissa Liu for their encouragement, and understanding that keep my inner stability for the past over ten years. I thank Professor Erik Eberhardt of The University of British Columbia for providing me a visiting opportunity to revise the manuscript. I thank Natural Resources Canada for access to the Canadian National Seismograph Network data archive, the seismograms used in this study can be obtained from the Canadian National Seismograph Network at http://www.earthquakescanada.nrcan.gc.ca/stndon/CNSN-RNSC/ .